\documentclass[12pt]{llncs}
\usepackage{amsmath,amssymb,fullpage,graphicx,mathptmx}

\begin{document}
\title{Effective closed subshifts in 1D can be\\ implemented in 2D}
\author{Bruno Durand\inst{1},
Andrei Romashchenko\inst{1,2}, Alexander Shen\inst{1,2}}
\institute{LIF, CNRS \& Univ. de Provence, Marseille
\and On leave from the Institute for Information Transmission Problems, Moscow}

\maketitle

\thispagestyle{plain}

\begin{abstract}

In this paper we use fixed point tilings to answer a question posed by Michael Hochman and show that every one-dimensional effectively closed subshift can be implemented by a local rule in two dimensions. The proof uses the fixed-point construction of an aperiodic tile set and its extensions.

\end{abstract}

\section{Introduction}

Let $A$ be a finite set (\emph{alphabet}); its elements are called \emph{letters}.
By $A$-\emph{configuration} we mean a mapping $C\colon \mathbb{Z}^2\to A$.  In geometric terms: a cell with coordinates $(i,j)$ contains letter $C(i,j)$.

A \emph{local rule} is defined by a positive integer $M$ and a list of prohibited $(M\times M)$-\emph{patterns} ($M\times M$ squares filled by letters). A configuration $C$ \emph{satisfies} a local rule $R$ if none of the patterns listed in $R$ appears in $C$.

Let $A$ and $B$ be two alphabets and let $\pi\colon A\to B$ be any mapping. Then every $A$-configuration can be transformed into a $B$-configuration (its homomorphic image) by applying $\pi$ to each letter.  Assume that the local rule $R$ for $A$-configurations and mapping $\pi$ are chosen in such a way that local rule $R$ prohibits patterns where letters $a$ and $a'$ with $\pi(a)\ne\pi(a')$ are vertical neighbors. This guarantees that every $A$-configuration that satisfies $R$ has an image where vertically aligned $B$-letters are the same. Then for each $B$-configuration in the image every vertical line carries one single letter of $B$. So we can say that $\pi$ maps a $2$-dimensional $A$-configuration satisfying the local rule $R$ to a $1$-dimensional $B$-configuration.

Thus, for every $A$, $B$, local rule $R$ and $\pi$ (with described properties) we get a subset $L(A,B,R,\pi)$ of $B^\mathbb{Z}$ (i.e., $L(A,B,R,\pi)$ is the set of $\pi$-images of all $A$-configurations that satisfy the local rule $R$). The following result (theorem~\ref{main}) characterizes the subsets that can be obtained in this way.

Consider a product topology in $B^\mathbb{Z}$. Base open sets of this topology are \emph{intervals}. Each interval is obtained by fixing letters in finitely many places (other places may contain arbitrary letters). Each interval is therefore a finite object and we can define \emph{effectively open} subsets of $B^\mathbb{Z}$ as unions of (computably) enumerable families of intervals. An \emph{effectively closed} set is a complement of an effectively open one. A \emph{subshift} is a subset of $B^\mathbb{Z}$ that is closed and invariant under left and right shifts. We are mostly interested in subshifts that are not only closed but effectively closed sets.

\begin{theorem}
	\label{main}
The subset $L(A,B,R,\pi)$ is an effectively closed subshift. For every effectively closed subshift $S\subset \mathbb{B}^Z$ one can find $A$, $R$, and $\pi$ such that $S=L(A,B,R,\pi)$.
\end{theorem}

The first part of the statement is easy. The set $L(A,B,R,\pi)$ is evidently shift invariant; it remains to show that it is effectively closed. The set of all $A$-configurations that satisfy $R$ is a closed subset of a compact space and therefore is a compact space itself. The mapping of $A$-configurations into $B$-configurations is continuous. Therefore the set $L(A,B,R,\pi)$ is compact (as a continuous image of a compact set). This argument can be effectivized in a standard way. A $B$-string is declared \emph{bad} if it cannot appear in the $\pi$-image of any $A$-configuration that satisfies $R$.  The set of all bad strings is enumerable and $L(A,B,R,\pi)$ is the set of all bi-infinite sequences that have no bad factors.

The reverse implication is more difficult and is the main subject of this paper. It cannot be proven easily since it implies the classical result of Berger~\cite{berger}: the existence of a local rule that makes all configurations aperiodic. Indeed, it is easy to construct an effectively closed subshift  $S$ that has no periodic points; if it is represented as $L(A,B,R,\pi)$, then local rule $R$ has no periodic configurations (configurations that have two independent period vectors); indeed, those configurations have a horizontal period vector.

So it is natural to expect a proof of theorem~\ref{main} to be obtained by modifying one of the existing constructions of an aperiodic local rule. It is indeed the case: we use the fixed-point construction described in~\cite{firstpart}. We do not repeat this construction (assuming that the reader is familiar with that paper or has it at hand) and explain only the modifications that are needed in our case. This is done in sections~\ref{sec:idea}--\ref{sec:degenerate}; in the rest of this section we survey some other steps in the same direction.

M.~Hochman~\cite{hochman} proved a similar result for 3D implementations of 1D subshifts (and, in general, $(k+2)$-dimensional implementation of $k$-dimensional subshifts) and asked whether a stronger statement is true where 3D is replaced by 2D. 

As we have mentioned, it is indeed true and can be achieved by the technique of fixed point self-similar tilings. The detailed exposition of this technique and its applications, including an answer to Hochman's question, is given in our paper~\cite{long}. Since this paper contains many other results (most boring are related to error-prone tile sets), we think that a self-contained (modulo~\cite{firstpart}) exposition could be useful for readers that are primarily interested in this result, and provide such an exposition in the current paper.

In fact, the fixed point construction of algorithms and machines is an old and well known tool (used, e.g., for Kleene's fixed point theorem and von Neumann's self-reproducing automata) that goes back to self-referential paradoxes and G\"odel's incompleteness theorem. One may only wonder why it was not used in 1960s to construct an aperiodic tile set. In a context of hierarchical constructions in the plane this technique was used by P.~G\'acs in a much more complicated situation (see~\cite{gacs2d}); however, G\'acs did not bother to mention explicitly that this technique can be applied to construct aperiodic tile sets. 

Fixed point tilings are not the only tool that can be used to implement subshifts. In~\cite{dls} a more classical (Berger--Robinson style) construction of an aperiodic tile set is modified in several ways to implement one specific shift: the family of bi-infinite bit sequences  $\omega$ such that all sufficiently long substrings $x$ of $\omega$ have complexity greater than $\alpha |x|$ or at least $\omega$ can be cut into two pieces (left- and right-infinite) that have this property. (Here $\alpha$ is some constant less than $1$, and $|x|$ stands for the length of $x$.)  In fact, the construction used there is fairly general and can be applied to any enumerable set $F$ of forbidden substrings: one may implement a shift that consists of bi-infinite sequences that have no substrings in $F$ or at least can be cut into two parts with this property. Recently N.~Aubrun and M.~Sablik  found a more ingenious construction that is free of this problem (splitting into two parts) and therefore provides another proof of theorem~\ref{main} (see~\cite{sablik}).

The authors thank their LIF colleagues, especially E.~Jeandel who pointed out that their result answers a question posed in~\cite{hochman}.

\section{The idea of the construction}
\label{sec:idea}

We do not refer explicitly to our paper~\cite{firstpart} but use the notions and constructions from that paper freely. In that paper we used local rules of special type (each letter was called a tile and has four colors at its sides; the local rule says that colors of the neighbor tile should match). In fact, any local rule can be reduced to this type by extending the alphabet; however, we do not need to worry about this since we construct a local rule and may restrict ourselves to tilings.

We superimpose two layers in our tiling. One of the layers contains $B$-letters; the local rule guarantees that each vertical line carries one $B$-letter. (Vertical neighbors should be identical.) For simplicity we assume that $B=\{0,1\}$, so $B$-letters are just bits, but this is not really important for the argument.

The second layer contains an aperiodic tile set constructed in a way similar to~\cite{firstpart}.  Then some rules are used to organize the interaction between the layers; computations in the second layer are fed with the data from the first layer and check that the first layer does not contain any forbidden string.

Indeed, the macro-tiles (at every level) in our construction contain some computation used to guarantee their behavior as building blocks for the next level.  Could we run this computation in parallel with some other one that enumerates bad patterns and terminates the computation (creating a violation of the rules) if a bad pattern appears?

\smallskip
This idea immediately faces evident problems:

\begin{itemize}

\item The computation performed in macro-tiles (in~\cite{firstpart})  was limited in time and space (and we need unlimited computations since we have infinitely many forbidden substrings and no limit on the computational resources used to enumerate them).

\item Computations on high levels do not have access to bit sequence they need to check: the bits that go through these macro-tiles are ``deep in the subconscious'', since macro-tiles operate on the level of their sons (cells of the computation that are macro-tiles of the previous level), not individual bits.

\item Even if every macro-tile checks all the bits that go through it (in some mysterious way), a ``degenerate case'' could happen where an infinite vertical line is not crossed by any macro-tile. Imagine a tile that is a left-most son of a father macro-tile who in its turn is the left-most son of its father and so on (see Fig.~\ref{degenerate.mps}). They fill the right half-plane; the left half-plane is filled in a symmetric way, and the vertical dividing line between then is not crossed by any tile. Then, if each macro-tile takes care of forbidden substrings inside its zone (bits that cross this macro-tile), some substrings (that cross the dividing line) remain unchecked.

\end{itemize}

These problems are discussed in the following sections one after another; we apologize if the description of them seemed to be quite informal and vague and hope that they would become more clear when their solution is discussed.

\section{Variable zoom factor}
\label{sec:variable}

In our previous construction the macro-tiles of all levels were of the same size: each of them contained $N\times N$ macro-tiles of the previous level for some constant zoom factor $N$. Now it is not enough any more, since we need to host arbitrarily long computations in high-level macro-tiles. So we need an increasing sequence of zoom factors $N_0,N_1,N_2,\ldots$; macro-tiles of the first level are blocks of $N_0\times N_0$ tiles; macro-tiles of the second level are blocks of $N_1\times N_1$ macro-tiles of level $1$ (and have size of $N_0N_1\times N_0N_1$ if measured in individual tiles).  In general, macro-tiles of level $k$ are made of $N_{k-1}\times N_{k-1}$ macro-tiles of level $k-1$ and have side $N_0N_1\ldots  N_{k-1}$ measured in individual tiles.

However, all the macro-tiles (of different levels) carry the same program in their computation zone. The difference between their behavior is caused by the data: each macro-tiles ``knows'' its level (consciously, as a sequence of bits on its tape). Then this level $k$ may be used to compute $N_k$ which is then used as a modulus for coordinates in the father macro-tile. (Such a coordinate is a number between $0$ and $N_k-1$, and the addition is performed modulo $N_k$.)

Of course, we need to ensure that this information is correct. Two properties are required: (1) all macro-tiles of the same level have the same idea about their level, and  (2) these ideas are consistent between levels (each father is one level higher than its sons). The first is easy to achieve: the level should be a part of the side macro-color and should match in neighbor tiles. (In fact an explicit check that brothers have the same idea about their levels is not really needed, since the first property follows from the second one: since all tiles on the  level zero ``know'' their level correctly, by induction we conclude that macro-tiles of all levels have correct information about their levels.)

To achieve the second property (consistency between level information consciously known to a father and its sons) is also easy, though we need some  construction. It goes as follows: each macro-tile knows its place in the father, so it knows whether the father should keep some bits of his level information in that macro-tile. If yes, the macro-tile checks that this information is correct.  Each macro-tile checks only one bit of the level information, but with brothers' help they check all the bits.\footnote{%
   People sitting on the stadium during the football match and holding color sheets to create a slogan for their team can check the correctness of the slogan by looking at the scheme and knowing their row and seat coordinates; each person checks one pixel, but in cooperation they check the entire slogan.}

There is one more thing we need to take care of: the level information should fit into the tiles (and the computation needed to compute $N_k$ knowing $k$ should also fit into level $k$ tile). This means that $\log k$, $\log N_k$ and the time needed to compute $N_k$ from $k$ should be much less than $N_{k-1}$ (since the computation zone is some fraction of $N_{k-1}$).  So $N_k$ should not grow too slow (say, $N_k=\log k$ is too slow), should not grow too fast (say, $N_k=2^{N_{k-1}}$ is too fast) and should not be too difficult to compute. However, these restriction still leave a lot of room for us: e.g., $N_k$ can be proportional to $\sqrt{k}$, to $k$, to $2^k$, or $2^{2^k}$, or $2^{2^{2^k}}$ (any fixed height is OK). Recall that computation deals with binary encodings of $k$ and $N_k$ and normally is polynomial in their lengths.

In this way we are now able to embed computations of increasing sizes into the macro-tiles. Now we have to explain which data these computations would get and how the communication between levels is organized. 

\section{Conscious and subconscious bits}

The problem of communication between levels can be explained using the following metaphor. Imagine you are a macro-tile; then you have some program, and process the data according to the program; you ``blow up'' (i.e., your interior cannot be correctly tiled) if some inconsistency in the data is found. This program makes you perform as one cell in the next-level brain (in the computation zone of the father macro-tile), but you do not worry about it: you just perform the program. At the same time each cell of yourself in fact is a son macro-tile, and elementary operations of this cell (the relation between signals on its sides) are in fact performed by a lower-level computation. But this computation is your ``sub-conscious'', you do not have direct access to its data, though the correct functioning of the cells of your brain is guaranteed by the programs running in your sons.

Please do not took this metaphor too seriously and keep in mind that the time axis of the computations is just a vertical axis on the plane; configurations are static and do not change with time. However, it could be useful while thinking about problems of inter-level communication.

Let us decide that for each macro-tile all the bits (of the bit sequence that needs to be checked) that cross this macro-tile form its \emph{responsibility zone}. Moreover, one of the bits of this zone may be \emph{delegated} to the macro-tile, and in this case the macro-tile consciously knows this bit (is \emph{responsible} for this bit). The choice of this bit depends on the vertical position of the macro-tile in its father. 

More technically, recall that a macro-tile of level $k$ is a square whose side is $L_k=N_0\cdot N_1\cdot \ldots\cdot N_{k-1}$, so there are $L_k$ bits of the sequence that intersect this macro-tile. We delegate each of these bits to one of the macro-tiles it intersects. Note that every macro-tile of the next level is made of $N_k\times N_k$ macro-tiles of level $k$. We assume that $N_k$ is much bigger than $L_k$ (more about choice of $N_k$ later); this guarantees that there are enough macro-tiles of level $k$ (in the next level macro-tile) to serve all bits that intersect them. Let us decide that $i$th macro-tile of level $k$ (from bottom to top) in a $(k+1)$-level macro-tile knows $i$th bit (from the left) in its zone. Since $N_k$ is greater than $L_k$, we leave some unused space in each macro-tile of level $k+1$: many macro-tiles of level $k$  are not responsible for any bit, but this does not create any problems.

\begin{figure}[h]
        $$
\includegraphics[scale=0.8]{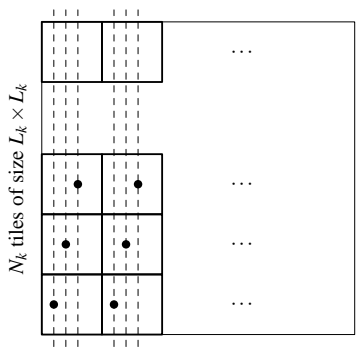}
        $$
\caption{Bit delegation: bits assigned to vertical lines are distributed between $k$-level macro-tile (according to their positions in the father macro-tile of level $k+1$).}
\label{fpt.8.mps}
\end{figure}

This is our plan; however, we need a mechanism that ensures that the delegated bits are indeed represented correctly (are equal to the corresponding bits ``on the ground'', in the sequence that forms the first level of our construction). This is done in the hierarchical way: since every bit is delegated to macro-tiles of all levels, it is enough to ensure that the ideas about bit values are consistent between father and son. 

For this hierarchical check let us agree that every macro-tile not only knows its own delegated bit (or the fact that there is no delegated bit), but also knows the bit delegated to its father (if it exists) as well as father's coordinates (in the grandfather macro-tile). This is still an acceptable amount of information (for keeping father's coordinates we need to ensure that $\log N_{k+1}$, the size of father's coordinate, is much smaller that $N_{k-1}$). To make this information consistent, we ensure that 

\begin{itemize}

\item
the data about the father's coordinates and bits are the same among brothers;

\item
if a macro-tile has the same delegated bit as its father (this fact can be checked since a macro-tile knows its coordinates in the father and father's coordinates in the grandfather), these two bits coincide;

\item
if a macro-tile is in the place where its father keeps its delegated bit, the actual father's information is consistent with the information about what the father should have.
 
\end{itemize}

So the information transfer between levels is organized as follows: the macro-tile that has the same delegated bit as its father, non-deterministically guesses this fact and distributes the information about father's coordinates and bit among the brothers. Those of the brothers who are in the correct place, check that father indeed has correct information.

On the lowest level we have direct access to the bits of the sequence, so the tile that is above the correct bit can keep its value and transmit it together with (guessed) coordinates of its father macro-tile (in the grandfather's macro-tile) to all brothers, and some brothers are in the right place and may check these values against the bits in the computation zone of the father macro-tile.

This construction makes all bits present at all levels, but this is not enough for checking: we need to check not individual bits, but bit groups (against the list of forbidden substrings). To this end we need a special arrangement described in the next section.

\section{Checking bit groups}

Here the main idea is: each macro-tile checks some substring (bit group) that is very small compared to the size of this macro-tile. However, since the size of the computation zone grows infinitely as the level increases, this does not prevent  complicated checks (that may involve a long substring that appears very late in the enumeration  of the forbidden patterns) from happening.

The check is performed as follows:  we do some number of steps in the enumeration of forbidden patterns, and then check whether one of these patterns appears in the bit group under consideration (assigned to this macro-tile). The number of enumeration steps can be also rather small compared to the macro-tile size. 

We reserve also some time and space to check that all the patterns appeared during the enumeration are not substrings of the bit group under consideration. This is not a serious time/space overhead since substring search in the given bit group can be performed rather fast, and the size of the bit group and the number of enumeration steps are chosen small enough (having in mind this overhead).

Then in the limit any violation inside some macro-tile will be discovered (and only degenerate case problem remains: substrings that are not covered entirely by any tile). The degenerate case problem is considered in the next section; it this section it remains to explain how the groups of (neighbor) bits are made available to the computation and how they are assigned to macro-tiles.

Let us consider an infinite vertical stripe of macro-tiles of level $k$  that share the same $L_k=N_0\cdot\ldots \cdot N_{k-1}$ columns. Together, these macro-tiles keep in their memory all $L_k$  bits  of  their common zone of responsibility. Each of them perform a check for a small bit group (of length $l_k$, which increases extremely slowly with $k$ and in particular is much less than $N_{k-1}$). We need to distribute somehow these groups among macro-tiles of this infinite stripe. 

It can be done in many different ways. For example, let us agree that the starting point of the bit group  checked by a macro-tile is the vertical coordinate of this macro-tile in its father (if it is not too big; recall that $N_k \gg N_0 N_1\ldots N_{k-1}$). It remains to explain how  groups of (neighbor) bits are made available to the computational zones of the corresponding macro-tiles.

We do it in the same way as for delegated bits; the difference (and simplification) is that now we may use only two levels of hierarchy since all the bits are available in the previous level (and not only in the ``deep unconscious'', at the ground level). We require that this group and the coordinate that determines its position are again known to all the sons of the macro-tile where the group is checked. Then the sons  should ensure that  (1)~this information is consistent between brothers; (2)~it is consistent with delegated bits where delegated bits are in the group, and (3)~it is consistent with the information in the macro-tile (father of these brothers) itself. Since $l_k$ is small, this is a small amount of information so there is no problem of its distribution between macro-tiles of the preceding level.

If a forbidden pattern belongs to a zone of responsibility of macro-tiles of arbitrarily high level, then this violation is be discovered  inside a macro-tile of some level, so the tiling of the plain cannot not exist. Only the degenerate case problem remains: so far we cannot catch forbidden substrings that are not covered entirely by any macro-tile. We deal with the degenerate case problem in the next section.

\section{Dealing with the degenerate case}
\label{sec:degenerate}

\begin{figure}[h]
        $$
\includegraphics[scale=0.8]{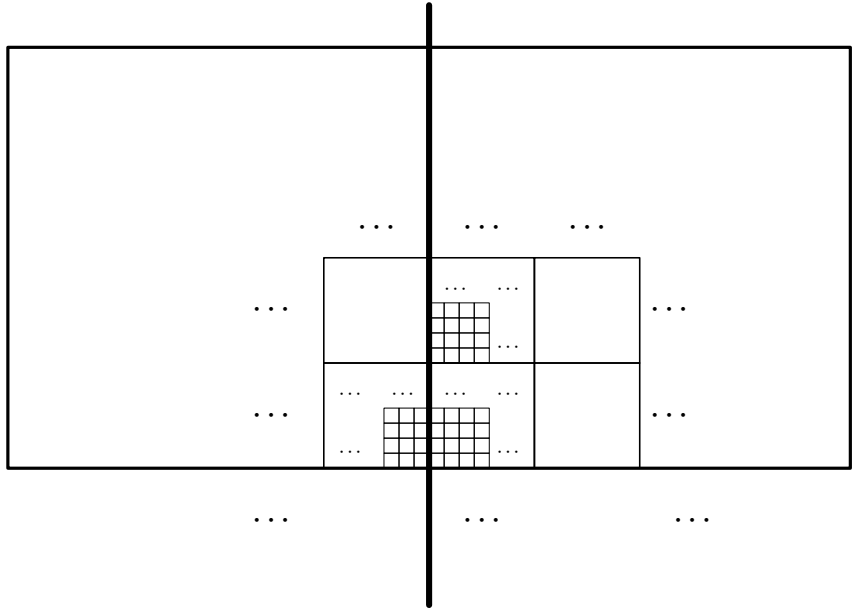}
        $$
\caption{Degenerate case}
\label{degenerate.mps}
\end{figure}

The problem we need to deal with: it can happen that one vertical line is not crossed by any macro-tile of any level (see Fig.~\ref{degenerate.mps}). In this case  some substrings are not covered entirely by any macro-tile, and we do not check them.  
After the problem is realized, the solution is not difficult.  We let every macro-tile check bit groups in its \emph{extended responsibility zone} that is three times wider and covers not only the macro-tile itself but also its left and right neighbors.

Now a  macro-tile of level $k$ is given a small bit group which is a substring of its extended responsibility zone (the width of the extended responsibility zone is $3L_k$; it is composed of the zones of responsibility of the macro-tile itself and two its neighbors). Respectively, a macro-tile of level $(k-1)$ keeps the information about three groups of bits instead of one: for its father, left uncle, and right uncle. This information should be consistent between brothers (since they have the same father and uncles). Moreover, it should be checked across the boundary between macro-tiles: if two macro-tiles $A$ and $B$ are neighbors but have different fathers ($B$'s father is $A$'s right uncle and $A$'s father is $B$'s left uncle), then they should compare the information they have (about bit groups checked by fathers of $A$ and $B$) and ensure it is consistent. For this we need to increase the amount of information kept in a macro-tile by a constant factor (a macro-tile keeps three bit groups instead of one, etc.), but this is still acceptable.

It is easy to see that now even in the degenerate case every substring is entirely in the extended responsibility zone of arbitrary large tiles, so all the forbidden patterns are checked everywhere.

\section{Final adjustments}

We finished our argument, but we was quite vague about the exact values of parameters saying only that some quantities should be much less than others. Now we need to check again the entire construction and see that the relations between parameters that were needed at different steps could be fulfilled together.

\medskip

Let us remind the parameters used at several steps of the construction: macro-tiles of level $k+1$ consist of $N_k\times N_k$ macro-tile of level $k$; thus, a $k$-level macro-tile consists of $L_k\times L_k$ tiles (of level $0$), where $L_k=N_0\cdot\ldots\cdot N_{k-1}$. Macro-tiles of level $k$ are responsible for checking bit blocks of length $l_k$ from their extended responsibility zone (of width $3L_k$). We have several constraints on the values of these parameters:

\begin{itemize}

\item $\log N_{k+1}\ll N_k$ and even $\log N_{k+2}\ll N_k$ since every macro-tile must be able to do simple arithmetic manipulations with its own coordinates in the father and with coordinates of the  father in the grandfather;

\item $N_k\gg L_k$ since we need enough sons of a macro-tile of level $k+1$ to keep all bits from its zone of responsibility (we use one macro-tile of level $k$ for each bit);

\item $l_{k}$ and even $l_{k+1}$ should be much less than $N_{k-1}$ since a macro-tile of level $k$ must contain in its computational zone the bit block of length $l_k$ assigned to itself and three bit blocks of length $l_{k+1}$ assigned to its father and two uncles (the left and right neighbors of the father);

\item a $k$-level macro-tile should enumerate in its computational zone several forbidden patterns and check whether any of them is a substring of the given (assigned to this macro-tile) $l_k$-bits block; the number of steps in this enumeration must be small compared to the size of the macro-tile; for example, let us agree that a macro-tile of level $k$ runs this enumeration for exactly $l_k$ steps;

\item the values $N_k$ and $l_k$ should be simple functions of $k$: we want to compute $l_k$ in time polynomial in $k$, and compute $N_k$ in time polynomial in $\log N_k$ (note that typically $N_k$ is much greater than $k$, so we cannot compute or even write down its binary representation in time polynomial in $k$).

\end{itemize}

With all these constraints we are still quite free in the choice of parameters. For example, we may let $N_{k}=2^{C 2^k}$ (for some large enough constant $C$) and $l_k=k$.  

\section{Final remarks}

One may also use essentially the same construction to implement $k$-dimensional effectively closed subshifts using $(k+1)$-dimensional subshifts of finite type.

How far can we go further? Can we implement  evert $k$-dimensional effectively closed subshifts by a tiling of the same dimension $k$? Another question (posed in \cite{hochman}): let us replace a finite alphabet by a Cantor space (with the standard topology); can we represent every $k$-dimensional effectively closed subshifts  over a Cantor space as a continuous image of the set of  tilings of dimension $k+1$ (for some finite tile set)?  E.~Jeandel noticed that the answers to the both questions are negative (this fact  is also a corollary of results from~\cite{dls} and \cite{rumush}).

\end{document}